\documentclass[aps,pra,twocolumn,showpacs,superscriptaddress,amsmath,amssymb,
10pt]{revtex4-1}
\usepackage{graphicx}
\graphicspath{{img/}}

\newcommand{\tr}{\mathop{\mathrm{tr}}}
\begin{document}
\title{Non-Markovian qubit decoherence during dispersive readout}

\author{Georg M. Reuther}
\author{Peter H\"anggi}
\affiliation{Institut f\"ur Physik, Universit\"at Augsburg,
Universit\"atsstra{\ss}e~1, D-86135 Augsburg, Germany}

\author{Sigmund Kohler}
\affiliation{Instituto de Ciencia de Materiales de Madrid, CSIC,
Cantoblanco, E-28049 Madrid, Spain}

\begin{abstract}
We study qubit decoherence under generalized dispersive readout,
i.e., we investigate a qubit coupled to a resonantly driven dissipative
harmonic oscillator.  We provide a complete picture
by allowing for arbitrarily large qubit-oscillator detuning
and by considering also a coupling to the square of the oscillator
coordinate, which is relevant for flux qubits.  Analytical results for
the decoherence time are obtained by a transformation of the
qubit-oscillator Hamiltonian to the dispersive frame and a subsequent
master equation treatment beyond the Markov limit.  We predict a
crossover from Markovian decay to a decay with Gaussian shape.  Our
results are corroborated by the numerical solution of the full
qubit-oscillator master equation in the original frame.
\end{abstract}

\date{\today}


\maketitle


\section{Introduction}

The final readout of the qubit state presents an essential part of any
quantum algorithm~\cite{Nielsen2000a}.  For solid state qubits, it is
typically realized by coupling the qubit to a harmonic oscillator such
that the oscillator frequency undergoes a shift whose sign depends on
the state of the qubit.  This shift can be probed by driving the
oscillator at its bare frequency, with the consequence that the phase
of the response provides information about the qubit
state~\cite{Chiorescu2004a, Wallraff2004a, Grajcar2004a,
Sillanpaa2005a, Lupascu2006a, Schuster2007a}.  Typically, one works
with a qubit-oscillator detuning that is slightly larger than the
respective mutual coupling while still being much smaller than the
qubit splitting~\cite{Chiorescu2004a, Wallraff2004a}. Then the
resulting frequency shift of the oscillator can be derived from the
Rabi Hamiltonian via a transformation to the so-called dispersive
frame~\cite{Blais2004a}.  For even stronger detuning, the
transformation becomes more involved, but nevertheless the sign of the
dispersive shift depends on the qubit state~\cite{Zueco2009b} and,
thus, qubit readout remains possible.  For an extreme detuning, such
that the oscillator frequency exceeds the qubit splitting by far, a
measurement protocol has been proposed by which one can reconstruct
information about coherent qubit oscillations from recorded
data~\cite{Reuther2011b}.  Even though similar readout is possible
by driving the qubit directly \cite{Reuther2009a, Reuther2011a}, the
oscillator plays a constructive role as band pass.

For driving the oscillator and for measuring its response, the setup
must be coupled to auxiliary electronic circuitry, which represents an
environment that eventually destroys the phase of the qubit.
Generally information about the qubit state can be obtained only at
the rate at which the qubit coherence decays~\cite{Clerk2010a}.  Thus
a quantitative understanding of qubit decoherence stemming from the
coupling to a resonantly driven oscillator is inevitable for the
design of dispersive readout schemes.  For sufficiently small
detuning, such that the rotating-wave approximation underlying the
Rabi Hamiltonian holds, the decoherence rate follows from an intuitive
consideration in which the shot noise of the cavity photons randomizes
the qubit phase~\cite{Blais2004a}.  This result will emerge as
limiting case of our more general picture.

For weakly dissipative quantum systems, Bloch-Redfield
theory~\cite{Redfield1957a, Blum1996a} represents a natural framework
for studying decoherence, in particular when memory effects are minor.
When qubit decoherence stems from a dissipative harmonic oscillator,
however, its naive application may significantly overestimate the
decoherence rate \cite{Thorwart2004a, Nesi2007a} because peaks in the
effective spectral density of the oscillator~\cite{Garg1985a,
Tian2002a, vanderWal2003a} cause non-Markovian behavior. This may in
particular be the case for the ultra-strong qubit-oscillator coupling
which marks a recent trend~\cite{Devoret2007a, Anappara2009a,
Ashhab2010a, Niemczyk2010a}.

Here we present a global picture of the qubit decoherence during
dispersive readout.  We consider both linear and quadratic
qubit-oscillator coupling as well as arbitrarily large detuning, while
we do not account for higher-order corrections to the dispersive
shift~\cite{Boissonneault2009a} and non-linearities in the oscillator
potential~\cite{Boissonneault2012a}.  While in Ref.~\cite{Tian2002a}
this problem was studied for an oscillator at thermal equilibrium, we
focus on the semiclassical regime, in which the oscillator
is governed by the driving.
In Sec.~\ref{sec:model}, we introduce our quantum master equation for
the full qubit-oscillator-bath model.  Section~\ref{sec:analytics} is
devoted to an analytical derivation of the dephasing time.  We develop
a picture in which the driven dissipative oscillator acts as a bath
which is eliminated in second-order perturbation theory but beyond a
Markov approximation.  In order to corroborate the resulting dephasing
time, we numerically solve in Sec.~\ref{sec:num} the quantum master
equation for the qubit and the oscillator in the original frame.

\section{Qubit-oscillator model}
\label{sec:model}

The qubit coupled to the oscillator is described by the Hamiltonian
\begin{equation}
\label{H0}
H_0 = \frac{\hbar\epsilon}{2}\sigma_z
      + \hbar g_1\sigma_x(a^\dagger+a)
      + \hbar g_2\sigma_x(a^\dagger+a)^2
      + \hbar\omega_0 a^\dagger a ,
\end{equation}
where $g_1$ is the strength of the dipole interaction which is linear in
the oscillator displacement $a^\dagger+a$. The coupling to
the square of the displacement with strength $g_2$ is
mainly relevant for flux qubits and can be controlled to some
extent~\cite{Bertet2005a,Bertet2005b}.

The system state can be probed via a coupling an external circuitry
which we model by the system-bath Hamiltonian
\begin{equation}
\label{Hbath}
H_\text{bath} = (a^\dagger+a) \sum_\nu c_\nu (b_\nu^\dagger+b_\nu)
+ \sum_\nu \hbar\omega_\nu b_\nu^\dagger b_\nu ,
\end{equation}
where $b_\nu$ is the annihilation operator of a bath mode $\nu$.
The influence of the bath is determined by the spectral
density $I(\omega) = 2\pi\sum_\nu c_\nu^2\delta(\omega-\omega_\nu)$,
which we assume to be ohmic, i.e., $I(\omega) = \gamma\omega/\omega_0$
with the oscillator damping rate $\gamma$.

An external ac driving corresponds to one particular bath mode being
in a highly excited coherent state.  It may be described as classical
oscillation, such that the oscillator experiences a driving force
$A\cos(\Omega t-\phi_0)$. This corresponds to the driving Hamiltonian
\begin{equation}
\label{HF}
H_F(t) = Ax \cos(\Omega t-\phi_0).
\end{equation}
For convenience, we have introduced the dimensionless position and momentum
operators $x = (a^\dagger+a)/\sqrt{2}$ and $p=i(a^\dagger-a)/\sqrt{2}$,
respectively.

Within the usual Born-Markov approximation for the oscillator (see
Appendix~\ref{app:ho}), one obtains the master equation
\begin{equation}
\label{Lfull}
\begin{split}
\dot R
= & -\frac{i}{\hbar}[H_0+H_F(t)]
    -\frac{i\gamma}{2}[x,[p, R]_+]
\\ &
-\frac{\gamma}{2}\coth\Big(\frac{\hbar\omega_0}{2k_BT}\Big)[x,[x, R]]
\end{split}
\end{equation}
for the joint density operator $R$ of the qubit and the oscillator.
It provides all numerical results presented below. In case of low
temperatures, $k_BT\ll\hbar\omega_0$, the hyperbolic cotangent assumes a
value close to unity. Notice that in contrast to a parametric driving,
the linearly coupled ac force does not affect the dissipative
terms~\cite{Kohler1997a}. Moreover, we assume that the qubit couples
only weakly to the oscillator, so that its influence on the
dissipative kernel of the oscillator is negligible.

\section{Analytical estimate of the qubit dephasing time}
\label{sec:analytics}

Extracting from the master equation \eqref{Lfull} an analytical
expression for the qubit decoherence time represents a formidable
task.  Thus, we have to rely on several approximations that
make use of the conditions under which dispersive readout may be
performed.  We start by a transformation to the dispersive frame which
yields a coupling to the square of the position coordinate of the
driven oscillator.  The relevant influence on the qubit is determined
by the corresponding auto correlation function which we evaluate in
the semiclassical limit.  It becomes stationary only after averaging
within a rotating-wave approximation over the initial phase $\phi_0$
of the driving.  Finally, the resulting non-Markovian master equation
for the qubit is solved for short times.

\subsection{Transformation to the dispersive frame}

For the discussion of dispersive readout, the dispersive picture of
the qubit-oscillator Hamiltonian $H_0$ has proven useful
\cite{Blais2004a, Clerk2010a, Reuther2011b}.  In order to capture also
very large qubit-oscillator detuning, we need to perform the
according transformation beyond rotating-wave approximation.  This
yields the effective qubit-oscillator Hamiltonian~\cite{Zueco2009b}
\begin{equation}
\label{Hdis}
\bar H_0 = \frac{\hbar\epsilon}{2}\sigma_z
   + \hbar(\lambda_\parallel \sigma_z +\lambda_\perp \sigma_x) x^2
   + \frac{\hbar\omega_0}{2}( x^2 + p^2 )
\end{equation}
with the coupling constants
\begin{align}
\label{lambda1}
\lambda_\parallel
={}& \frac{g_1^2}{\epsilon-\omega_0} + \frac{g_1^2}{\epsilon+\omega_0}
,
\\
\label{lambda2}
\lambda_\perp ={}& 2g_2 .
\end{align}
Thus, the coupling linear in the oscillator coordinate has turned into
a quadratic coupling with strength $\lambda_\parallel$, while
$\lambda_\perp$ has been introduced for a unified notation.
In correspondence to the orientation of the coupling operators on the
Bloch sphere, we refer to the coupling terms as ``longitudinal'' and
``transverse'', respectively.

If only energy-conserving terms in the qubit-oscillator coupling were
considered \cite{Blais2004a}, $\lambda_\parallel$ would be given by
only the first term of Eq.~\eqref{lambda1} and, thus, be $\propto
(\epsilon-\omega_0)^{-1}$.  By contrast, the second term of
Eq.~\eqref{lambda1} turns the frequency dependence into
$\lambda_\parallel \propto |\epsilon^2-\omega_0^2|^{-2}$.  This means
that for positive detuning, $\omega_0>\epsilon$, the counter-rotating
terms diminish the dispersive shift.  Since we will find that
decoherence grows with $\lambda_\parallel$, the coherence time is
larger than predicted within rotating-wave
approximation~\cite{Blais2004a}.

The interpretation of the effective interaction is that it shifts the
oscillator frequency by $\pm (\lambda_\parallel^2
+\lambda_\perp^2)^{1/2}$, where the sign depends on the state of the
qubit.  Therefore, probing the oscillator frequency provides
information about the latter. In this work we are interested in
the qubit decoherence that stems from this coupling.  When
transforming to the dispersive frame, the dissipative terms of the
master equation~\eqref{Lfull} acquire a contribution from qubit
operators \cite{Boissonneault2009b, Reuther2010a}, which, however, is
negligible as compared to the terms considered below.

\subsection{Driven oscillator as effective bath}

We now treat the oscillator as environment coupled to the qubit
coordinate $Y=\lambda_\parallel\sigma_z +\lambda_\perp\sigma_x$ via
the Hamiltonian $\bar H_\text{int} = \hbar Y \eta$ with the
environmental fluctuations $\eta$ determined by the operator $x^2$.
Its expectation value $\langle x^2\rangle$ yields a correction to the
qubit Hamiltonian of the order $\lambda_\parallel$.  Therefore, the
impact of $\langle x^2\rangle$ on the dissipative terms is already
beyond the order considered herein and can be omitted, such that the
relevant fluctuation reads
\begin{equation}
\eta = x^2 - \langle x^2\rangle .
\end{equation}
We assume weak dissipation such that the bath can be eliminated within
second order perturbation theory.  This is in accordance with our
observation of predominantly coherent time evolution; see the
numerical results in Sec.~\ref{sec:num}.  Then the dissipative part of
the master equation for the qubit density operator $\rho$ reads
\begin{equation}
\label{QME}
\begin{split}
\dot\rho
=
- \int_0^t dt\Big\{
  & S_{\eta\eta}(t,t') [Y,[Y(t-t'),\rho]] \\
+ & iA_{\eta\eta}(t,t')[Y,[Y(t-t'),\rho]_+] \Big\} .
\end{split}
\end{equation}
It is non-Markovian owing to its explicit time-dependence and the
corresponding lack of a semi-group property.
Here $[\ ,\ ]_+$ denotes the anti-commutator, while
\begin{equation}
\label{tildeY}
Y(t) = \lambda_\parallel\sigma_z
+ \lambda_\perp\{\sigma_x\cos(\epsilon t)-\sigma_y\sin(\epsilon t)\}
\end{equation}
is the qubit part of the coupling in the interaction picture.
$S_{\eta\eta}$ and $A_{\eta\eta}$ are the real part and
the imaginary part, respectively, of the effective bath correlation
function
\begin{equation}
C_{\eta\eta}(t,t')
= \langle \eta(t)\eta(t')\rangle
\equiv S_{\eta\eta}(t,t')+iA_{\eta\eta}(t,t') ,
\end{equation}
which is not time homogeneous due to the driving.
The evaluation of $S_{\eta\eta}$ in the limit in which
dispersive readout is performed is a cornerstone of our analytical
treatment.

Since we consider measurement schemes that rely on the response to
deterministic driving, the fluctuations are small so that we can
linearize in the oscillator position fluctuation $\delta x \equiv
x-\bar x(t)$ and work with the approximation
\begin{equation}
\eta(t) = 2\bar x(t) \delta x(t) .
\end{equation}
Then the auto correlation function of the effective bath coordinate
$\eta$ becomes
\begin{equation}
\label{Ceta0}
C_{\eta\eta}(t,t')
= 4 \bar x(t) \bar x(t') \, \langle \delta x(t) \delta x(t')\rangle ,
\end{equation}
where the term with the angular brackets is the position-position
correlation function $C_{\delta x\,\delta x}$ of the dissipative
harmonic oscillator.  Owing to the linearity of the oscillator's
equation of motion, it is independent of the driving and, thus,
stationary.

The response to the classical driving can be expressed in terms of the
oscillator Green's function $G(\omega)$, Eq.~\eqref{G0}.  This yields
\begin{equation}
\label{x(t)}
\bar x(t) = \sqrt{2\bar n} \cos(\Omega t-\phi_0-\phi) ,
\end{equation}
where $\bar n =\frac{1}{2}A^2|G(\omega)|^2$ is the mean cavity photon
number and $\phi$ is the phase of the Green's function, while $\phi_0$
is the unknown initial phase of the driving.  For a harmonic
oscillator in the weak coupling regime, the response to the
fluctuations of the external circuitry is conveniently computed with
the help of the quantum regression theorem~\cite{Lax1963a}, see
Appendix \ref{app:ho.qrt}.  For low temperatures,
$k_BT\ll\hbar\omega_0$, it can be approximated by
\begin{equation}
\label{S(t)}
S_{\delta x\,\delta x}(\tau)
= \frac{1}{2} e^{-\gamma \tau/2} \cos(\omega_0 \tau) ,
\end{equation}
where $\tau=t-t'$.
Equations \eqref{x(t)} and \eqref{S(t)} allow us to evaluate the
correlation function \eqref{Ceta0}.  After performing an average over
the initial phase $\phi_0$, we obtain for its symmetric
part the time homogeneous expression
\begin{equation}
\label{Ceta}
S_{\eta\eta}(\tau)
= 2\bar n e^{-\gamma \tau/2}\cos(\Omega \tau)\cos(\omega_0 \tau) .
\end{equation}
The phase average represents a rotating-wave approximation and is
possible since qubit decoherence and dissipation are much slower than
the coherent oscillator dynamics.
The correlation function \eqref{Ceta} possesses four resonance peaks
of width $\gamma$ at the frequencies $\omega = \pm\omega_0\pm\Omega$,
where for resonant driving, $\Omega=\omega_0$, the two central peaks
coincide at zero frequency.

The peaks of the spectral density correspond to long-time
correlations of the quantum noise that may lead to non-Markov effects.
Therefore, a treatment with a fully Markovian master equation is not
appropriate~\cite{Nesi2007a}.  We thus employ ideas that have been
used for studying decoherence due to $1/f$ noise~\cite{Breuer2003a,
Makhlin2004a, Falci2004a, Falci2005a, Matsuzaki2010a}.  We analyze the
longitudinal and the transverse dephasing separately by setting either
$g_1$ or $g_2$ to zero.

\subsection{Coherence decay for linear qubit-oscillator coupling}
\label{sec:puritydecay}

We assume that the qubit is initially in the state
$(|{\uparrow}\rangle + |{\downarrow}\rangle)/\sqrt{2}$, i.e., in a
coherent superposition of the eigenstates of the qubit Hamiltonian
$(\hbar\epsilon/2)\sigma_z$.  Then the corresponding off-diagonal
elements of the density matrix in the eigenbasis are both $1/2$ and
undergo an oscillatory decay, $\rho_{\uparrow\downarrow} =
\frac{1}{2}\exp\{-i\epsilon t-\Lambda(t)\}$,
where in the Markov limit, $\Lambda(t) = \Gamma t$.  It is
straightforward to demonstrate that then the purity $P \equiv \tr\rho^2$,
being our measure of coherence, evolves as
\begin{equation}
\label{P(t)}
P(t) = \frac{1}{2}[1+e^{-2\Lambda(t)}] \approx 1- \Lambda(t) .
\end{equation}
The approximation holds for short times at which the purity
still lies significantly above $1/2$.  A typical time evolution is
depicted in Fig.~\ref{fig:timeevolution} below.  It demonstrates that the
purity decay is indeed not necessarily a simple exponential, but may
have Gaussian shape.

The still unknown function $\Lambda(t)$ will be determined from the
master equation \eqref{QME} for the density matrix element
$\rho_{\uparrow\downarrow}$ at short times yielding for $g_2=0$
\begin{equation}
\label{dotLambda}
\dot\Lambda_\parallel(t)
= -\frac{\dot\rho_{\uparrow\downarrow}}{\rho_{\uparrow\downarrow}}
\equiv \Gamma_\parallel(t)
\end{equation}
with the time-dependent decoherence rate
\begin{align}
\label{Gamma-par}
\Gamma_\parallel(t)
={} & 4\lambda_\parallel^2 \int_0^t d\tau\, S_{\eta\eta}(\tau) .
\end{align}
The index $\parallel$ refers to the longitudinal qubit-oscillator
coupling in the dispersive Hamiltonian \eqref{Hdis}.  Notice that in
the original Hamiltonian \eqref{H0}, this coupling is transverse.
In the following, we evaluate this rate for resonant driving and weak
oscillator damping, $\gamma \ll \omega_0 = \Omega$.

Inserting the effective spectral density \eqref{Ceta} into
Eq.~\eqref{Gamma-par} yields
\begin{equation}
\label{Gamma-par2}
\Gamma_\parallel(t)
= \frac{8\bar n\lambda_\parallel^2}{\gamma}
  \big( 1 - e^{-\gamma t/2}\big) ,
\end{equation}
where we have neglected terms oscillating rapidly with frequency
$2\omega_0$.  By straightforward time integration, we obtain
\begin{align}
\label{Lambda(t)}
\Lambda_\parallel(t)
={}& \frac{8\bar n\lambda_\parallel^2}{\gamma^2}(\gamma t+2e^{-\gamma t/2}-2)
\\\approx{}&
\begin{cases}
  2 \bar n\lambda_\parallel^2 t^2 & \text{for $\gamma t \ll 1$},
  \\[2ex]
  8\bar n\lambda_\parallel^2 t/\gamma &\text{for $\gamma t \gg 1$}.
\label{Lambda(t)approx}
\end{cases}
\end{align}
Inserting this approximation into our ansatz for
$\rho_{\uparrow\downarrow}$ reveals that during an initial stage, the
coherence decays like a Gaussian $\rho_{\uparrow\downarrow}\sim
\exp(-t^2/T_G^2)$ with the time scale
\begin{equation}
\label{TG}
T_G = \frac{1}{\sqrt{2\bar n\lambda_\parallel^2}}.
\end{equation}
Thereafter, normal exponential decay $\rho_{\uparrow\downarrow}\sim
\exp(-t/T_M)$ sets in, where
\begin{equation}
\label{TM}
T_M = \frac{\gamma}{8\bar n \lambda_\parallel^2}.
\end{equation}
Since both approximations in Eq.~\eqref{Lambda(t)approx} are never
smaller than the exact expression, we connect the two limits by
choosing at each time the smaller one, i.e.,
\begin{equation}
\Lambda_\parallel(t) \approx \min(t^2/T_G^2, t/T_M) .
\end{equation}
This implies a crossover from Gaussian to Markovian decay at time
$t_c = 4/\gamma$.

As criterion for ``significant dephasing'', we use $\Lambda>1/4$,
which means that the off-diagonal matrix element
$\rho_{\uparrow\downarrow}$ has decayed by at least 22 percent.  For
larger values of $\Lambda$, the visibility of coherent oscillations is
already quite small.  Therefore, the relevant dephasing is Gaussian if
$t_c^2/T_G^2 > 1/4$ or, equivalently, $128\bar{n}\lambda_\parallel^2 >
\gamma^2$.  In the opposite case, the Gaussian stage can be ignored
and coherence fades away during time $T_M/4$.  In combination, this
yields the dephasing time
\begin{equation}
\label{Tpar}
T_\parallel^* \approx
\begin{cases}
    T_M/4 = \frac{\gamma}{32\bar n \lambda_\parallel^2}
    & \text{for $128\bar{n}\lambda_\parallel^2 < \gamma^2$} ,
\\[2ex]
    T_G/2 = \frac{1}{\sqrt{8\bar n\lambda_\parallel^2}}
    & \text{for $128\bar{n}\lambda_\parallel^2 > \gamma^2$} .
\end{cases}
\end{equation}
The first line holds for the Markovian behavior found for weak
coupling.  Notice that $\lambda_\parallel$ is an effective coupling
constant that becomes smaller with increasing detuning
$|\epsilon-\omega_0|$.  Thus, for small detuning (but still within the
dispersive limit) and for large photon number, we expect Gaussian
decay.

At this stage, it is interesting to establish a connection to
Refs.~\cite{Blais2004a, Clerk2010a}, where the fluctuation of the
cavity photon number leads to a fluctuating qubit splitting and, thus,
randomizes the qubit phase.  Then, for $|\lambda_\parallel|
\lesssim\gamma$ and $\lambda_\perp=0$, one finds that the off-diagonal
matrix elements of the density operator decay with a rate $\Gamma_\phi
=1/T_M = 8\lambda_\parallel^2 \bar n/\gamma$~\cite{Blais2004a,
Clerk2010a}, which is accordance with our result in the Markov limit.

\subsection{Coherence decay for quadratic qubit-oscillator coupling}
\label{sec:analytics-g2}

For $g_2\neq0$, the situation becomes considerably more complicated,
because the dissipative terms in the master equation \eqref{QME}
couple all density matrix elements to each other.  Therefore, one can
no longer obtain a closed first-order equation for
$\rho_{\uparrow\downarrow}$ such as Eq.~\eqref{dotLambda}.  For this
reason we attempt an analytical solution only in the Markovian
regime.  In doing so, we perform the time integral in Eq.~\eqref{QME}
until infinity such that we obtain a time-independent Bloch-Redfield
master equation.  The decoherence rate is conveniently
extracted from the equivalent equation of motion for the Bloch vector
$\vec s = \langle\vec\sigma\rangle$. By straightforward algebra the
latter emerges as $d\vec s/dt = M\vec s + \vec h$ with the dynamical
matrix
\begin{equation}
M=
\begin{pmatrix}
0 & -\epsilon & 0 \\
\epsilon & -2\Gamma_\perp & 0 \\
0 & 0 & -2\Gamma_\perp
\end{pmatrix}
\end{equation}
and the decay rate
\begin{equation}
\label{Gamma}
\Gamma_\perp = \lambda_2^2 S_{\eta\eta}(\epsilon)
= \frac{4\gamma\bar{n}\lambda_\perp^2}{\gamma^2+4\epsilon^2} .
\end{equation}
The inhomogeneity $\vec h$ stems from the second line of
Eq.~\eqref{QME} and determines the stationary state which is not
relevant in the present context.

We proceed by computing the eigenvalues of $M$ to lowest order in the
dissipative terms, which yields $-2\Gamma_\perp$ and $\pm
i\epsilon-\Gamma_\perp$ \cite{Fonseca2004a}.  The latter
correspond to the decaying oscillations of
$\rho_{\uparrow\downarrow}$, which reveals that the transverse
decoherence is determined by $\Lambda_\perp = \Gamma_\perp t$.  Thus,
we obtain in the Markov limit the time scale $T_M = 1/\Gamma_\perp$
and, thus, the dephasing time
\begin{equation}
\label{Tperp}
T_\perp^* = \frac{1}{4\Gamma_\perp} =
\frac{\gamma^2+4\epsilon^2}{16\gamma\bar{n}\lambda_\perp^2} .
\end{equation}
This result holds under two conditions.  First, it is required that
the time-integration in the master equation~\eqref{QME} can be
extended to infinity, which is possible if the decay time of the
effective bath correlation function~\eqref{Ceta} is much shorter than
the dephasing time, $1/\gamma\ll T_\perp^*$, which means $16
\bar{n}\lambda_\perp^2\ll \gamma^2+4\epsilon^2$. Second, the qubit
frequency must be within the oscillator linewidth, i.e.,
$\epsilon\lesssim\gamma$, because otherwise the oscillator would
shield the qubit from the external circuitry.  Then higher-order
processes may dominate while the master equation~\eqref{QME} is of
only second order.  The latter condition is also essential for an
application that we have in mind, namely time-dependent dispersive
qubit readout via a high-frequency oscillator~\cite{Reuther2011b}.

\section{Numerical determination of the qubit dephasing time}
\label{sec:num}

The numerical computation of the dephasing time $T^*$ from the full
master equation~\eqref{Lfull} is possible only in a restricted
parameter regime for various reasons.  First, at resonant driving, the
stationary state of the oscillator has mean photon number $\bar n =
A^2/2\gamma^2$.  The assumption of the oscillator being in its
semiclassical limit is fulfilled only for $\bar n \gg 1$ or
equivalently $A\gg\gamma$.  Second, our considerations in
Sec.~\ref{sec:model} require that the oscillator reaches its
stationary state during a time much shorter than the qubit dephasing
time, i.e., for $1/\gamma \ll T_\parallel, T_\perp$.  Finally,
the stationary photon number is limited by computation time, which
grows with the number of oscillator Fock states needed for numerical
convergence.  While we find a good agreement of the numerical and the
analytical results already for $\bar n\gtrsim 5$, the natural
expectation is that the agreement even increases with the mean photon
number, because then the oscillator becomes more classical.

We start with the oscillator in the coherent state that corresponds to
the stationary classical solution in the absence of the qubit.  For
the qubit itself, we use as initial state the superposition
$(|{\uparrow}\rangle + |{\downarrow}\rangle)/\sqrt{2}$.  The
dissipative time evolution of both the qubit and the oscillator is
obtained by numerical integration of the full master equation~\eqref{Lfull}
with the original Hamiltonian~\eqref{H0}.  Thus we implicitly also
test the validity of the dispersive picture in the presence of a heat
bath.

The central quantity of our numerical study is the time evolution of
the purity $P(t) = \tr\rho^2(t)$ from which we determine the
dephasing time $T^*$ by the criterion $P(T^*) =
\frac{1}{2}(1+e^{-1/2})$, i.e., $\Lambda(T^*)=1/4$ as above.
Moreover, we use $P(t)$ to decide whether decoherence decays like a
simple exponential or like a Gaussian.  A formal procedure for the
distinction is fitting $P(t)$ for short times to the ansatz $P(t)
= 1-a_Mt-a_G^2t^2$.  The decay is mainly Markovian or mainly
Gaussian depending on which rate $a_M$ or $a_G$ is larger and
thus dominates.

\subsection{Linear qubit-oscillator coupling $g_1$}

\begin{figure}
\centerline{\includegraphics{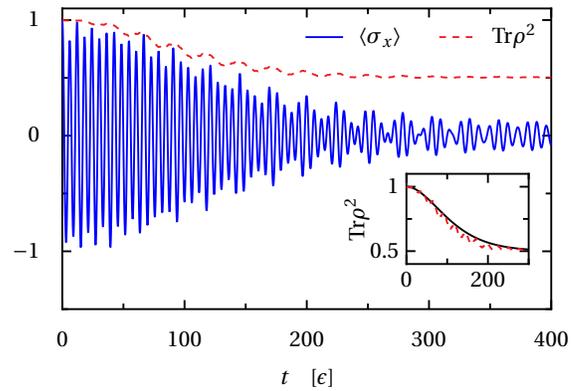}}
\caption{(Color online) Typical time evolution of the qubit operator
$\langle\sigma_x\rangle$ (solid line) and the corresponding purity
(dashed) for $\Omega = \omega_0 = 0.8\epsilon$, $g_1=0.02\epsilon$,
$\gamma =0.02\epsilon$, and driving amplitude $A=0.06\epsilon$ such
that the stationary photon number is $\bar n = 4.5$.  Inset: Purity decay
shown in the main panel (dashed) compared to the decay given by
Eq.~\eqref{P(t)} together with Eq.~\eqref{Lambda(t)} (solid line).
}
\label{fig:timeevolution}
\end{figure}
Figure~\ref{fig:timeevolution} depicts the time evolution of the qubit
expectation value $\langle\sigma_x\rangle$ which exhibits decaying
oscillations with frequency $\epsilon$.  The parameters correspond to
an intermediate regime between the Gaussian and the Markovian
dynamics, as is visible in the inset.

\begin{figure}
\centerline{\includegraphics{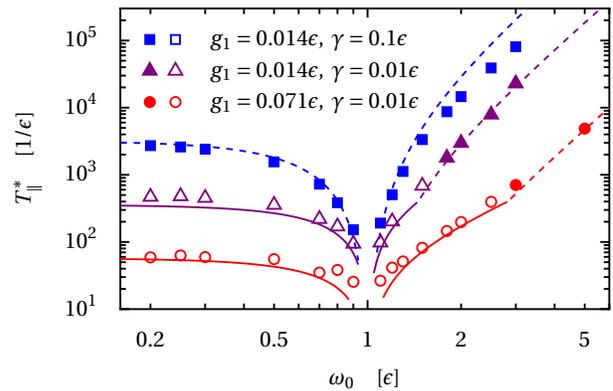}}
\caption{(Color online) Dephasing time for purely linear
qubit-oscillator coupling ($g_2=0$), resonant driving, $\Omega =
\omega_0$, and oscillator damping $\gamma=0.02\epsilon$.  The
amplitude $A=0.07\epsilon$ corresponds to the mean photon number $\bar
n = 6.125$.
Filled symbols and dashed lines refer to predominantly Markovian decay,
while for Gaussian decay, stroked symbols and solid lines are used.
}
\label{fig:Tlong}
\end{figure}
In Fig.~\ref{fig:Tlong}, we compare the decay time $T_\parallel^*$ with the
analytical result~\eqref{Tpar} for various values of the oscillator
damping and the qubit-oscillator coupling $g_1$ as function of the
oscillator frequency.  Whether Markovian or Gaussian decay dominates
is visualized by filled and stroked symbols, respectively. We
have skipped the regime very close to resonance, $|\lambda_\parallel|
\lesssim 5|\omega_0 -\epsilon|$, since there the dispersive
Hamiltonian~\eqref{Hdis} is not valid and so far no dispersive readout
protocol has been proposed.

Our prediction for regimes with non-Markovian decay is confirmed by
the numerical solution rather well.  Notice however, that in the
crossover regime, our formal criterion for Markovian decay provides a
unique answer, even though the respective other decay may already contribute
significantly.  The agreement of the numerically found border with our
prediction corroborates as well the crossover time $t_c$ conjectured
above.  Concerning the values of $T_\parallel^*$, we observe a good overall
agreement with the tendency that the analytical result slightly
underestimates $T_\parallel^*$.  In the regime of Gaussian decay,
a Markov approximation would yield a significantly smaller coherence
time.  This means that, interestingly enough, the qubit stays coherent
for a longer time than is expected from Bloch-Redfield theory.  For large
oscillator frequency, $\omega_0\gg\epsilon$, also the predicted
behavior $T_\parallel^* \propto \lambda_\parallel^{-2} \propto
|\epsilon^2-\omega_0^2|^2$ is confirmed.  This substantiates the
relevance of the counter-rotating correction to the dispersive
shift~\cite{Zueco2009b}.

\subsection{Quadratic qubit-oscillator coupling $g_2$}

We proceed as above, but consider the coupling to the square of
the oscillator coordinate in the Hamiltonian~\eqref{H0}, while setting
$g_1=0$. Even though the linear coupling $g_1$ can be controlled to
some extent~\cite{Bertet2005a, Bertet2005b}, it is probably hard to
turn it off completely. Still our choice has relevance in the limit of
large detuning in which the effective dispersive coupling
$\lambda_\parallel$ becomes rather small, see Eq.~\eqref{lambda1}.
Then for realistic values of $g_2$ for flux qubits, a protocol for
recording coherent time evolution has been proposed
\cite{Reuther2011b}.  A necessary condition for this is an oscillator
bandwidth of the order of the qubit splitting, such that the
oscillator does not filter out the information about the coherent
qubit dynamics. Therefore, we will choose  an oscillator with the
rather large frequency $\Omega=\omega_0 = 5\epsilon$ and with damping
up to $\gamma=\epsilon$.

\begin{figure}
\centerline{\includegraphics{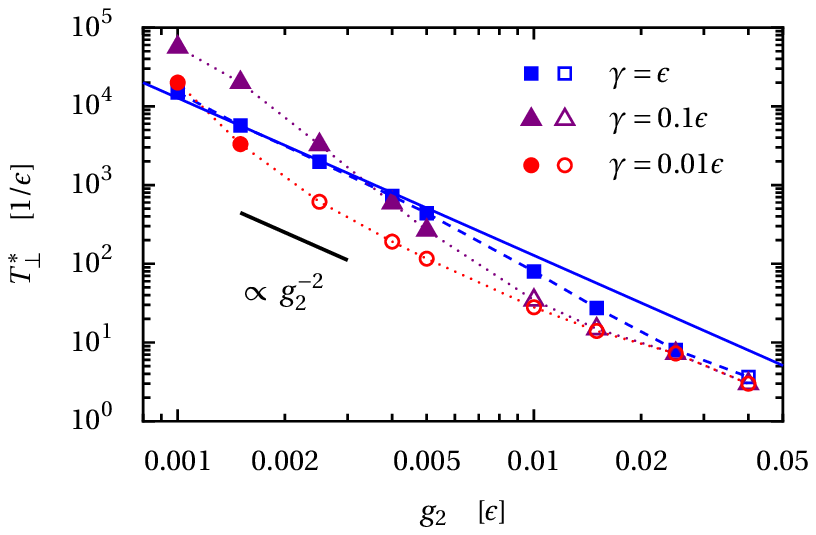}}
\caption{(Color online) Dephasing time for purely quadratic
qubit-oscillator coupling ($g_1=0$), resonant driving at large
frequency, $\Omega = \omega_0 = 5\epsilon$, and various values of the
oscillator damping $\gamma$.  The driving amplitude is $A=3.5\gamma$,
such that always $\bar n=6.125$.  Filled symbols mark Markovian decay,
while stroked symbols refer to Gaussian shape.  The solid line depicts
the value obtained for $\gamma=\epsilon$ in the Markov limit.  The
corresponding numerical values are connected by a dashed line which
serves as guide to the eye.
}
\label{fig:Ttrans}
\end{figure}
Figure~\ref{fig:Ttrans} shows the numerically obtained coherence times
and whether the decay is predominantly Gaussian or Markovian.  For the
large oscillator damping $\gamma=\epsilon$, the conditions for the
validity of the (Markovian) Bloch-Redfield equation stated at the end
of Sec.~\ref{sec:analytics-g2} hold.  Then we observe a good agreement of
the numerically obtained $T_\perp^*$ and Eq.~\eqref{Tperp}.  There
seems a slight systematic deviation for large values of $g_2$.  In
this limit, however, the numerical results are not very precise,
because the purity falls off already during the first few oscillation
periods.  This rapid decay hinders a precise numerical determination
of the dephasing rate.  Thus, the agreement is still within the
numerical precision.

For small oscillator damping, $\gamma\ll\epsilon$, Bloch-Redfield
theory is not applicable, as discussed above.  Moreover, the short
time evolution of the purity is already too complex for a
comprehensive prediction of the decoherence time.  Nevertheless our
numerical solution provides some hint on the decoherence process.  We
find that the crossover from simple exponential decay to decay with a
Gaussian shape occurs at smaller values of the qubit-oscillator
coupling.  For small $g_2$, the qubit stays coherent slightly longer,
while for large $g_2$, coherence decays a bit faster as compared to
the case $\gamma=\epsilon$.  In both regimes, the $\gamma$ dependence
of $T_\perp^*$ is weak.  This disproofs the Markovian theory for
$\gamma\ll\epsilon$, because the latter predicts $T_M \propto
1/\gamma$, which stays in contrast to our numerical result.

\section{Discussion and conclusions}

A qubit undergoing dispersive readout, i.e., one that is coupled to a
resonantly driven dissipative harmonic oscillator, experiences
decoherence from a rather exotic effective environment. The latter's
main properties stem from the small oscillator linewidth, the strong
driving, and a coupling coordinate that does not commute with the
qubit Hamiltonian.  Nevertheless, it has been possible to analytically
obtain essential and concise information about the decoherence
process.  Our approach is based on a transformation to the dispersive
frame, which turns the linear coupling into phase noise. In doing so
it is crucial to perform the transformation beyond rotating-wave
approximation, in particular for studying the far detuned case where
the non-rotating corrections are of the same order as the
rotating-wave terms.  For the subsequent analytical treatment, we have
derived the dephasing time within our picture of an effective spectral
density provided by the driven harmonic oscillator in the
semiclassical limit.  At the same time it has turned out that a peaked
spectral density induces a generally non-Markovian dissipative
dynamics.

As a main finding of this work, we have pointed out that the
decoherence process happens in two stages. In the beginning, the
purity decays like a Gaussian, while subsequently, Markovian decay
sets in.  If the qubit-oscillator coupling is strong or if the
oscillator is strongly driven, the major part of the coherence decays
already during the first stage such that the relevant dynamics
possesses a Gaussian time profile.  Thus, with the trend towards
ultra-strong coupling between a qubit and a harmonic
mode~\cite{Devoret2007a, Anappara2009a, Niemczyk2010a, Ashhab2010a},
Gaussian decay should become increasingly relevant.  In the opposite
limit of weak coupling, the first stage reduces the qubit coherence
only by a small amount, rendering the relevant decoherence Markovian.
The dephasing times in the two regimes exhibit distinct parameter
dependencies, which we have determined analytically.  Remarkably, in
the Gaussian regime, the coherence time may be significantly longer
than what one would expect from an extrapolation of the Markovian
result.

For a numerical description of the dynamics, we have solved the full
Bloch-Redfield master equation for the qubit coupled to the driven
oscillator in the frame of the original Hamiltonian.  This has allowed
us to obtain numerical results that are fully independent of the
analytical treatment.  They have confirmed our predictions for the
partially non-Markovian purity decay for the case of linear
qubit-oscillator coupling. For quadratic coupling, the
decoherence process is more involved. Nevertheless, it has been
possible to obtain an analytical expression for the decoherence rate
in the Markovian limit.  Beyond this limit, our numerical solution
indicates that decoherence is non-Markovian provided that the
oscillator dissipation is very weak.

Finally, we are convinced that our results on the decoherence induced
by a resonantly driven oscillator will support the design of future
experiments with dispersive readout and its generalizations.

\begin{acknowledgments}
This work was supported by DFG through the collaborative research
center SFB 631 and by the Spanish Ministry of Economy and
Competitiveness through grant No.\ MAT2011-24331.
\end{acknowledgments}

\appendix
\section{The driven dissipative harmonic oscillator}
\label{app:ho}

\subsection{Markovian master equation}

In the absence of the qubit and for sufficiently weak oscillator-bath
coupling, the dissipative dynamics of the oscillator is well described
within Bloch-Redfield theory~\cite{Redfield1957a, Blum1996a}.  Since
additive driving of the form \eqref{HF} does not influence the
dissipative terms~\cite{Kohler1997a}, it is possible to use the master
equation for the undriven dissipative harmonic oscillator and to
simply add the time-dependent Hamiltonian \eqref{HF}, such that
\begin{equation}
\label{Losc}
\begin{split}
\dot\rho_\text{osc}
= & -i\omega_0 [a^\dagger a,\rho_\text{osc}]
-\frac{i}{\hbar}[H_F(t),\rho_\text{osc}]
\\
& -\frac{i\gamma}{2}[x,[p,\rho_\text{osc}]_+] - \frac{\gamma}{2}
  \coth\Big(\frac{\hbar\omega_0}{2k_BT}\Big)[x,[x,\rho_\text{osc}]].
\end{split}
\end{equation}

\subsection{Average position}

Owing to the linearity of the quantum Langevin equation for the
dissipative harmonic oscillator, its position expectation value
$\bar x$ obeys the classical equation of motion
\begin{equation}
\ddot{\bar x} +\gamma\dot{\bar x}+\omega_0^2 \bar x = \omega_0F(t) .
\end{equation}
The response $\bar x(t)$ to the driving is most conveniently obtained
by a time convolution with the Green's function $G(t)$, where
\begin{equation}
\label{G0}
G(\omega)
= \frac{\omega_0}{-\omega^2-i\gamma\omega+\omega_0^2},
\end{equation}
and the inhomogeneity $F(t)$.  For $F(t)=A\cos(\Omega t-\phi_0)$, the
solution reads
\begin{equation}
\label{app:xbar}
\bar x(t) = |G(\Omega)| A\cos(\Omega t-\phi-\phi_0) ,
\end{equation}
where the phase shift $\phi$ is the argument of the Green's function.
The corresponding semiclassical state is a coherent state with mean
photon number $\bar n = \frac{1}{2}A^2|G(\Omega)|^2$.


\subsection{Position correlation function}
\label{app:ho.qrt}

In the Markovian limit for the undriven dissipative harmonic
oscillator implied in the master equation \eqref{Losc} in the absence
of $H_F$, the equilibrium position auto-correlation function
\begin{equation}
S_{\delta x\,\delta x}(t)
= \frac{1}{2}\langle\delta [x(t),\delta x(0)]_+\rangle
\end{equation}
can be computed by help of the quantum regression theorem
\cite{Lax1963a}.  It essentially states that $S_{\delta x\,\delta
x}$ obeys the same equation of motion as the average position.  Thus,
\begin{equation}
\ddot S_{\delta x\,\delta x} +\gamma\dot S_{\delta x\,\delta x}
+\omega_0^2 S_{\delta x\,\delta x}
 = 0 .
\end{equation}
The initial value is
\begin{equation}
S_{\delta x\,\delta x}\big|_{t=0} = \frac{1}{2}
  \coth\Big(\frac{\hbar\omega_0}{2k_BT}\Big),
\end{equation}
while for symmetric ordering, its time derivative at $t=0$ vanishes.
Thus, we can express the solution as Fourier integral of the Green's
function \eqref{G0} which we evaluate via Cauchy's theorem.  In the
weak damping limit considered herein, $\gamma\ll\omega_0$, we
neglect the dissipation induced frequency shift and continue with the
approximation
\begin{equation}
S_{\delta x\,\delta x}(t) = \frac{1}{2}
\coth\Big(\frac{\hbar\omega_0}{2k_BT}\Big) e^{-\gamma t/2}\cos(\omega_0 t).
\end{equation}

%

\end{document}